\documentclass[%
 preprint,
 superscriptaddress,
 amsmath,amssymb,
 aps,
]{revtex4-2}
\usepackage{float}
\usepackage{graphicx}% Include figure files
\usepackage{dcolumn}% Align table columns on decimal point
\usepackage{bm}% bold math
\begin{document}

\preprint{APS/123-QED}

\title{Unconsciousness reconfigures modular brain network dynamics}

\author{Sofía Morena del Pozo}
\affiliation{Departamento de Física, Facultad de Ciencias Exactas y Naturales, Universidad de Buenos Aires, Buenos Aires, Argentina}
\affiliation{Instituto de Física de Buenos Aires (IFIBA), CONICET, Buenos Aires, Argentina}
\author{Helmut Laufs}
\affiliation{Department of Neurology, Christian Albrechts University, Kiel, Germany}
\author{Vincent Bonhomme}
\affiliation{Anesthesia and Intensive Care Laboratory, GIGA-Consciousness, GIGA Institute, University of Liège, Liège, Belgium}
\affiliation{University Department of Anesthesia and Intensive Care Medicine, Centre Hospitalier Régional de la Citadelle, Liège, Belgium}
\affiliation{Department of Anesthesia and Intensive Care Medicine, Centre Hospitalier Universitaire de Liège (CHU Liège), Liège, Belgium}
\author{Steven Laureys}
\affiliation{Coma Science Group, GIGA Consciousness, University of Liège, Liège,Belgium}
\author{Pablo Balenzuela}
\affiliation{Departamento de Física, Facultad de Ciencias Exactas y Naturales, Universidad de Buenos Aires, Buenos Aires, Argentina}
\affiliation{Instituto de Física de Buenos Aires (IFIBA), CONICET, Buenos Aires, Argentina}
\author{Enzo Tagliazucchi}
\affiliation{Departamento de Física, Facultad de Ciencias Exactas y Naturales, Universidad de Buenos Aires, Buenos Aires, Argentina}
\affiliation{Instituto de Física de Buenos Aires (IFIBA), CONICET, Buenos Aires, Argentina}
%\maketitle

\begin{abstract}
The dynamic core hypothesis posits that consciousness is correlated with simultaneously integrated and differentiated assemblies of transiently synchronized brain regions. We represented time-dependent functional interactions using dynamic brain networks, and assessed the integrity of the dynamic core by means of the flexibility and largest multilayer module of these networks. As a first step, we constrained parameter selection using a newly developed benchmark for module detection in heterogeneous temporal networks. Next, we applied a multilayer modularity maximization algorithm to dynamic brain networks computed from functional magnetic resonance imaging (fMRI) data acquired during deep sleep and under propofol anesthesia. We found that unconsciousness reconfigured network flexibility and reduced the size of the largest spatiotemporal module, which we identified with the dynamic core. Our results present a first characterization of modular brain network dynamics during states of unconsciousness measured with fMRI, adding support to the dynamic core hypothesis of human consciousness.
 \end{abstract}


\flushbottom
\maketitle

\thispagestyle{empty}

%%%%%%%%%%%%%%%%%%%%%%%%%%%%%%%%%%%%%%%%%%%%%%%%%%%%%%%%%%%%
%%%%%%%%%%%%%%%%%%%%%%%%%%%%%%%%%%%%%%%%%%%%%%%%%%%%%%%%%%%%
\section*{Introduction}

The dynamic nature of consciousness is already manifest in William James' 1890 metaphor of the ``stream of consciousness'' \cite{james2007principles}. According to James, consciousness ``does not appear to itself chopped up in bits'' but instead flows and reconfigures itself over time, much like the stream of a river. This metaphor has informed several scientific hypothesis concerning the relationship between conscious experience and neural processes in the human brain, perhaps the most influential being Edelman and Tononi's dynamic core hypothesis \cite{edelman2000reentry,tononi1998consciousness, edelman2008universe}. According to this hypothesis, consciousness must be understood as a process that unfolds over time (the ``dynamic core'') comprising an ever-changing network of regions that exchange information over relatively short time spans. This position is in sharp contrast with the proposal that activity within certain specific and localized neuroanatomical structures comprise the neural basis for consciousness \cite{rees2002neural}. 

\begin{figure}[t]
	\centering
	\includegraphics[width=\linewidth]{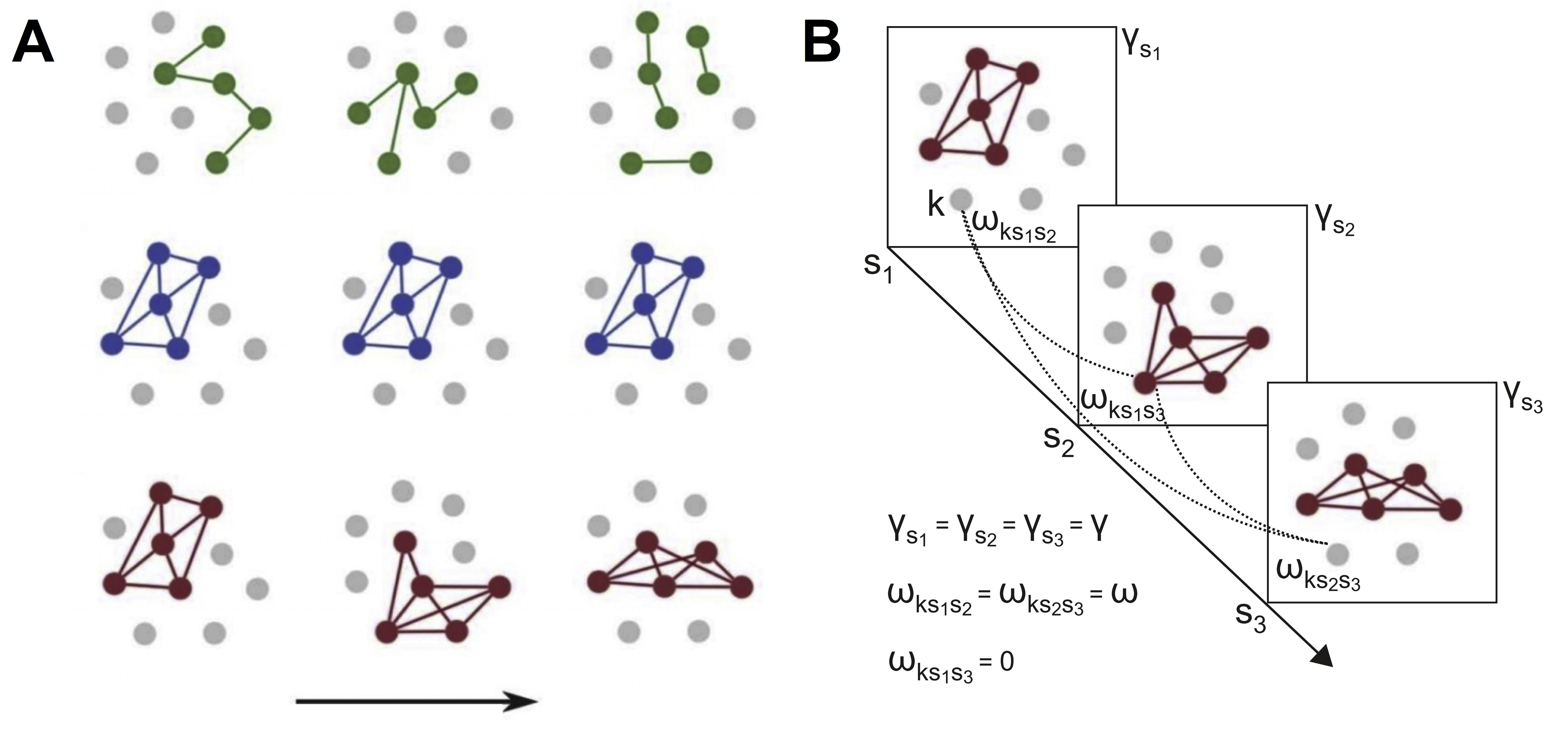}

 	\caption{A. Schematic representation of the dynamic core hypothesis. Each circle stands for a neuron or groups of neurons, and the links between them indicate transient synchronization. The first row depicts a sequence of highly differentiated but segregated states (green assemblies). The second row depicts a sequence of highly integrated states lacking differentiation (blue assemblies). Finally, the third row depicts a highly integrated dynamic core shifting through  several different configurations (red assemblies). B. These temporally evolving assemblies can be represented as modules unfolding in a multilayer network (layers $S_1$, $S_2$, $S_3$),  with inter-layer couplings for node $k$ given by $\omega_{kS_1S_2}$, $\omega_{kS_1S_3}$, $\omega_{kS_2S_3}$, and resolution parameters $\gamma_{S_1}$, $\gamma_{S_2}$ ,$\gamma_{S3}$. In this example, all interlayer couplings and resolution parameters are equal, and interlayer connections only appear between consecutive slices (e.g. $\omega_{KS1S3}=0$). Figure adapted from Cavanna et al, 2018 \cite{cavanna2018dynamic}}

	\label{figure1}
\end{figure}

The properties of the dynamic core are shaped by the tension between two observations concerning the nature of subjective experience \cite{tononi2016integrated}. On the one hand, conscious experience is extraordinarily rich, in the sense that admits countless different contents and variations. On the other hand, each individual conscious experience is integrated and cannot be decomposed into more elementary independent subexperiences (e.g. the senses of smell, sight and hearing are never experienced separately, but always fused together into the ongoing contents of consciousness). Thus, the dynamic core should present a very large number of possible configurations, corresponding to the multitude of available conscious experiences. However, these configurations must be constrained to represent highly integrated brain states. While a sequence of randomly connected brain regions is a highly differentiated process, each individual state in the sequence lacks integration (Fig. \ref{figure1}A, first row). Conversely, a stable sequence of fully connected regions lacks differentiation (Fig. \ref{figure1}A, second row). In between these two examples, the dynamic core consists of a sequence exploring an ample repertoire of highly integrated brain states (Fig. \ref{figure1}A, third row). 

The presence of simultaneously integrated and differentiated dynamics can be captured by different metrics of neural complexity \cite{tegmark2016improved}. For instance, unconscious brain states have been characterized by their entropy \cite{Carhart-Harris2014}, algorithmic complexity \cite{schartner2015complexity} and long-range information integration \cite{king2013information}. Furthermore, the persistence and complexity of activity evoked by external stimulation depends on the level of consciousness for different brain states such as deep sleep, general anesthesia, and disorders of consciousness \cite{Casali2013}. These changes are consistent with the identification of subcritical dynamics as a signature of unconsciousness, which results in stable collective neural behaviour that is incompatible with the rapid reconfiguration of metastable states postulated by the dynamic core hypothesis \cite{tagliazucchi2016large, tagliazucchi2017signatures, bocaccio2019avalanche}.

Perhaps the most straightforward method to assess the dynamic core hypothesis is also the most underexplored to date. The theory of complex networks provides a framework to directly assess the presence of integration and segregation in neuroimaging data acquired during different states of consciousness \cite{sporns2013network}. A sequence of brain states can be represented as a multilayer network, with each layer encoding transient functional interactions between brain regions during a given time period, while the dynamic core can be represented as a time-dependent module evolving in this network (Fig. \ref{figure1}B) \cite{muldoon2016network}. Over the last years, modularity maximization algorithms have been applied to multilayer networks to reveal the rapid and transient structure of whole-brain dynamic networks. These dynamics are sensitive to factors such as cognitive performance \cite{braun2015dynamic, telesford2016detection, bassett2011dynamic, pedersen2018multilayer} and neuropsychiatric impairments \cite{gifford2020resting}. In spite of these advances, however, the relationship between consciousness and the modular structure of multilayer brain networks remains to be investigated.

To clarify this relationship, we constructed multilayer connectivity networks from functional magnetic resonance imaging (fMRI) recordings acquired during the different stages of human non-rapid eye movement (NREM) sleep, and under the effect of propofol, a general anesthetic which increases inhibitory neurotransmission. Our main purpose was to obtain the time-dependent modular structure of these networks using the multilayer Louvain algorithm, a method with multiple free parameters related to the connectivity strength between temporal layers, and the characteristic size of the detected modules (Fig. \ref{figure1}B) \cite{mucha2010community}. Previous uses of this algorithm either employed an \emph{ad-hoc} choice of parameter values, or performed an exhaustive exploration of parameter space \cite{muldoon2016network,bassett2018nature, braun2015dynamic, telesford2016detection, bassett2011dynamic, pedersen2018multilayer, gifford2020resting}. We avoided these suboptimal solutions by introducing a new benchmark for the detection of modules in time-dependent networks with scale-free degree and module size distributions, adapted from a benchmark developed for static networks \cite{lancichinetti2008benchmark}. After parameter selection, we applied the multilayer Louvain algorithm to obtain the time-dependent modular structure of fMRI functional connectivity networks. We focused our analysis on two metrics related to the dynamics of modules over time: the flexibility (number of times a node switched its module assignment over time) and the size of the largest multilayer module, which we identified with the dynamic core.

%%%%%%%%%%%%%%%%%%%%%%%%%%%%%%%%%%%%%%%%%%%%%%%%%%%%%%%%%%%%
%%%%%%%%%%%%%%%%%%%%%%%%%%%%%%%%%%%%%%%%%%%%%%%%%%%%%%%%%%%%
\section*{Materials and methods}

\subsection*{fMRI data sets}

We analyzed two different fMRI data sets comprising healthy subjects, both during conscious wakefulness and during states of reduced consciousness.\\

\textbf{\textit{Propofol-induced loss of consciousness:}} 18 healthy volunteers (14 females, mean age: 22.8 years, range: 19-31) were scanned with fMRI during wakefulness (W), propofol sedation (S), propofol-induced loss of consciousness (LOC) and during the recovery of wakefulness (R) (written informed consent, approval by the local ethics committee). The Ramsay scale was used to determine the level of consciousness. Subjects were asked to squeeze strongly one hand, then their level of consciousness was evaluated based on strength and rapidity of their response. Resting state fMRI was acquired in all four states. The typical scan duration was half an hour in each condition. Functional images were acquired on a 3 Tesla Siemens Allegra scanner (Siemens AG, Munich, Germany; Echo Planar Imaging sequence using 32 slices; repetition time = 2460 ms, echo time = 40 ms, field of view = 220 mm, voxel size = 3.45$\times$3.45$\times$3 mm$^3$, and matrix size = 64$\times$64$\times$32). Previous publications based on this dataset can be consulted for further details \cite{boveroux2010breakdown}\\

\textbf{\textit{Human NREM sleep:}} fMRI data from 63 subjects acquired during wakefulness (W), 27 during N1 sleep, 33 during N2 sleep and 17 during N3 sleep (written informed consent, approval by the local ethics committee) in Frankfurt (Germany). EEG via a cap (modified BrainCapMR, Easycap, Herrsching, Germany) was recorded continuously during fMRI acquisition (1505 volumes of T2$^*$-weighted echo planar images, TR/TE = 2080 ms/30 ms, matrix 64$\times$64, voxel size 3$\times$3$\times$2 mm$^3$, distance factor 50$\%$; FOV 192 mm$^2$) with a 3 T Siemens Trio (Erlangen, Germany). An optimized polysomnographic setting was employed (chin and tibial EMG, ECG, EOG recorded bipolarly [sampling rate 5 kHz, low pass filter 1 kHz] with 30 EEG channels recorded with FCz as the reference [sampling rate 5 kHz, low pass filter 250 Hz].  Scalp potentials measured with EEG allowed the classification of sleep into 4 stages (wakefulness, N1, N2 and N3 sleep) according to the American Academy of Sleep Medicine (AASM) rules \cite{berry2012aasm}. Pulse oxymetry and respiration were recorded via sensors from the Trio [sampling rate 50 Hz]) and MR scanner compatible devices (BrainAmp MR+, BrainAmpExG; Brain Products, Gilching, Germany), facilitating sleep scoring during fMRI acquisition. Further details on data acquisition and preprocessing can be found in a previous publication \cite{tagliazucchi2014decoding}. \\

\subsection*{Module detection in multilayer networks}

We consider a multilayer network with adjacency matrix given by $A_{ijs}$, were $i,j$ index the network node and $s$ indexes the layer, which is here interpreted as a temporal dimension. Given a certain partition, its multilayer modularity (Q) is computed as,

\begin{equation}
Q= \frac{1}{2\mu}\sum_{ijrs}\left[\left(A_{ijs}-\gamma_s\frac{k_{is}k_{js}}{2m_s}\delta_{sr}\right)+\delta_{ij}\omega_{jsr}\right]\delta(g_{is},g_{jr})
\end{equation}

where $k_{js}=\sum_{i}A_{ijs}$, $\mu = \frac{1}{2}\sum_{jr}(k_{jr}+\sum_{s}\omega_{jrs})$, $m_s=\sum_j k_{js}$ and $\delta(g_{is},g_{jr})$ equals 1 if node $i$ of layer $s$ belongs to the same module as node $j$ of layer $r$. $\gamma_s$ is the resolution parameter for layer $s$, and $\omega_{jrs}$ represents the interlayer connectivity of node $j$ between layers $r$ and $s$. Here we consider the same $\gamma_s$ for all layers, and $\omega_{jrs} \neq 0$ only if $r$ and $s$ are consecutive layers, furthermore, all non-zero entries of  $\omega_{jrs}$ are equal.

Based on this target modularity function, we applied a generalized multilayer version of the Louvain algorithm to detect and track modules over time (\url{http://netwiki.amath.unc.edu/GenLouvain/GenLouvain}). The output of this algorithm was a 2D matrix $G_{it}$ presenting the module assignment of nodes ($i$, rows) across time ($t$, columns).

\subsection*{Benchmark for time-dependent module detection}

%Module detection algorithms can be evaluated using artificial networks with known modular structure, known as benchmarks. The simplest benchmark for module detection algorithms is the planted L-partition model \citep{plantedmodel}. This model consists of a graph with $n=g.L$ nodes partitioned into $L$ groups with $g$ nodes each. Nodes of the same group are connected to each other with a  probability $p_{in}$ and nodes of different groups are connected with probability $p_{out}$. The average degree of the nodes is computed as $<k>=p_{in}(g-1)+p_{out}g(L-1)$. If $p_{in}>p_{out}$ the density of intra-modular links is larger than the density of inter-modular links, resulting in detectable modular structure \citep{fortunato}. This model was adapted by Girvan and Newman to create one of the most widespread benchmarks \citep{GNbenchmark}. 

% Far from being homogeneous, however, real networks are complex, with modules of different sizes and scale-free (i.e. power-law) degree distributions. 

 To approach the problem of constructing a benchmark for module detection in complex time-dependent networks, we first reproduced a benchmark for static complex modular networks introduced by Lancichinetti et al. \cite{lancichinetti2008benchmark}.
 
 This benchmark assumes that both the degree and the module size distributions follow power laws, with exponents $\gamma$ and $\beta$, respectively.  The number of nodes is $N$ and the average degree is $<k>$. A mixing parameter is defined as $ \mu = \frac{\# \text{intra-modular links}}{\# \text{ total of links}} $, so each node shares a fraction  $1-\mu$ of its links with nodes of its own module and a fraction $\mu$ with the other nodes of the network.
 %Each node shares a fraction  $1-\mu$ of its links with nodes of its own module and a fraction $\mu$ with the other nodes of the network, $k_{in}=(1-\mu)k$, where $ \mu = \frac{\# \text{intra-modular links}}{\# \text{ total of links}} $ is the mixing parameter.
 Both the degrees of nodes, $k$, and the size of the modules, $s$, are constrained between predefined minimum and maximum values. The values of $k_{min}$ and $k_{max}$ are chosen such that the average degree is $<k>$ and $s_{min}$ and $s_{max}$ are chosen so that any node can belong to at least one module, $s_{min}>k_{min}(1-\mu)$ and $s_{max}>k_{max}(1-\mu)$. Further details can be found in Lancichinetti et al. (2018)\cite{lancichinetti2008benchmark}.

\begin{figure}[t]
	\centering
	\includegraphics[width=\linewidth]{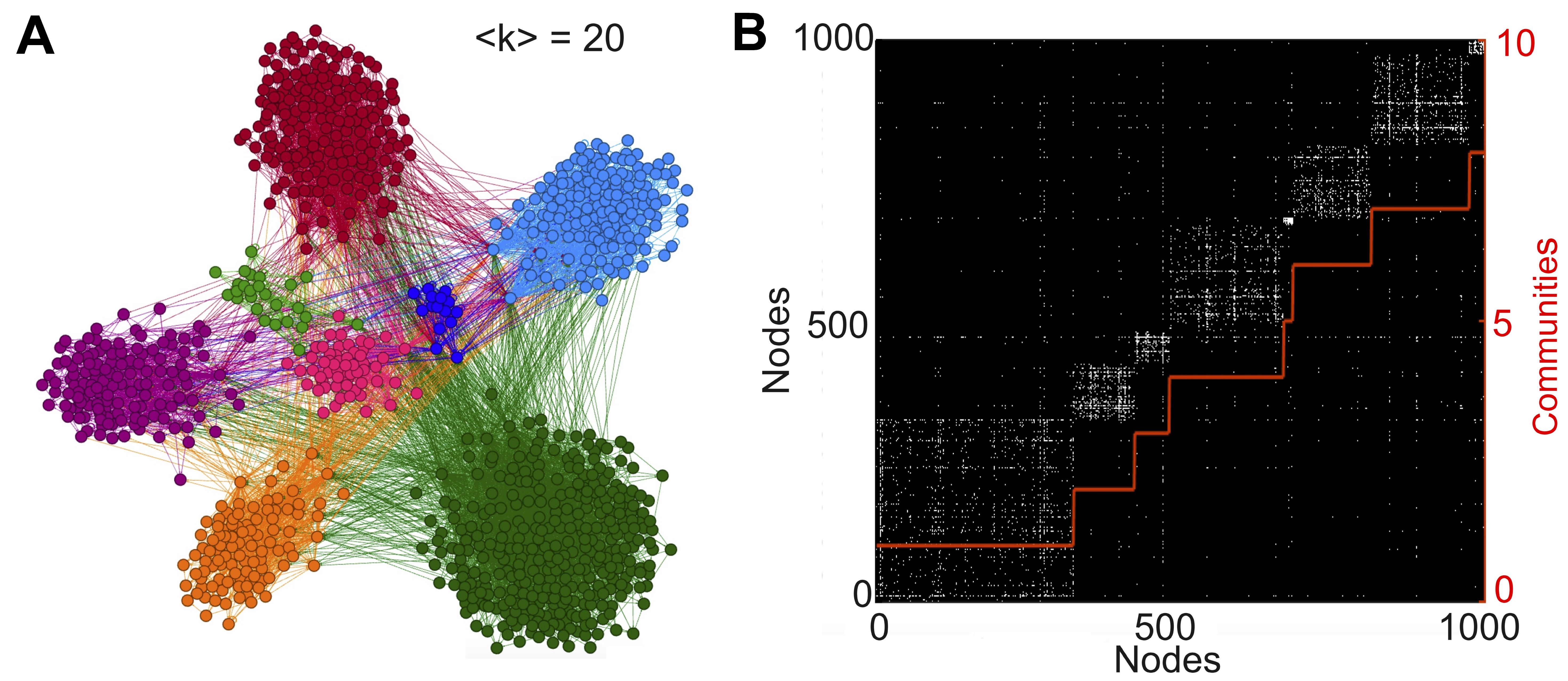}

 	\caption{(A) Graph representation of a network constructed following the benchmark introduced by Lancichinetti and colleagues with  $N=1000$, $\mu=0.07$, $\gamma=2$, $\beta=1$ and $<k>=20$ ($k_{min}=5$ and $k_{max}=360$). Modules obtained using the Louvain algorithm are indicated using different colors. (B) Matrix representation of the same network, with nodes sorted according to their module membership.}

	\label{figure2}
\end{figure}

We reproduced this static benchmark with $N=1000$, $\mu=0.07$, $\gamma=2$, $\beta=1$ and $<k>=20$ ($k_{min}=5$ and $k_{max}=360$) as a starting point to develop a time-dependent model for heterogeneous modular networks (see Fig. \ref{figure2}).

The proposed temporal evolution for our benchmark consisted of two different dynamic processes adapted from Granell et al., 2015 \cite{granell2015benchmark}: dynamics of merge-split and grow-shrink of modules. The combination of these two processes allowed us to represent the most frequent behaviours seen in the dynamics of real modular systems.

\subsubsection*{Null model based on division of communities: merge-split }\label{merge_split}

% Merges and splits of modules can occur in many real-world systems; for instance, groups of friends or colleagues can divide into smaller and less connected groups over time, or become suddenly polarized by discussions in social media. Another example arises from the dynamics of voting by political parties, with frequent merges and splits that reflect the forming and dissolving of coalitions. 

\begin{figure}[H]
	\centering
	\includegraphics[width=0.8\linewidth]{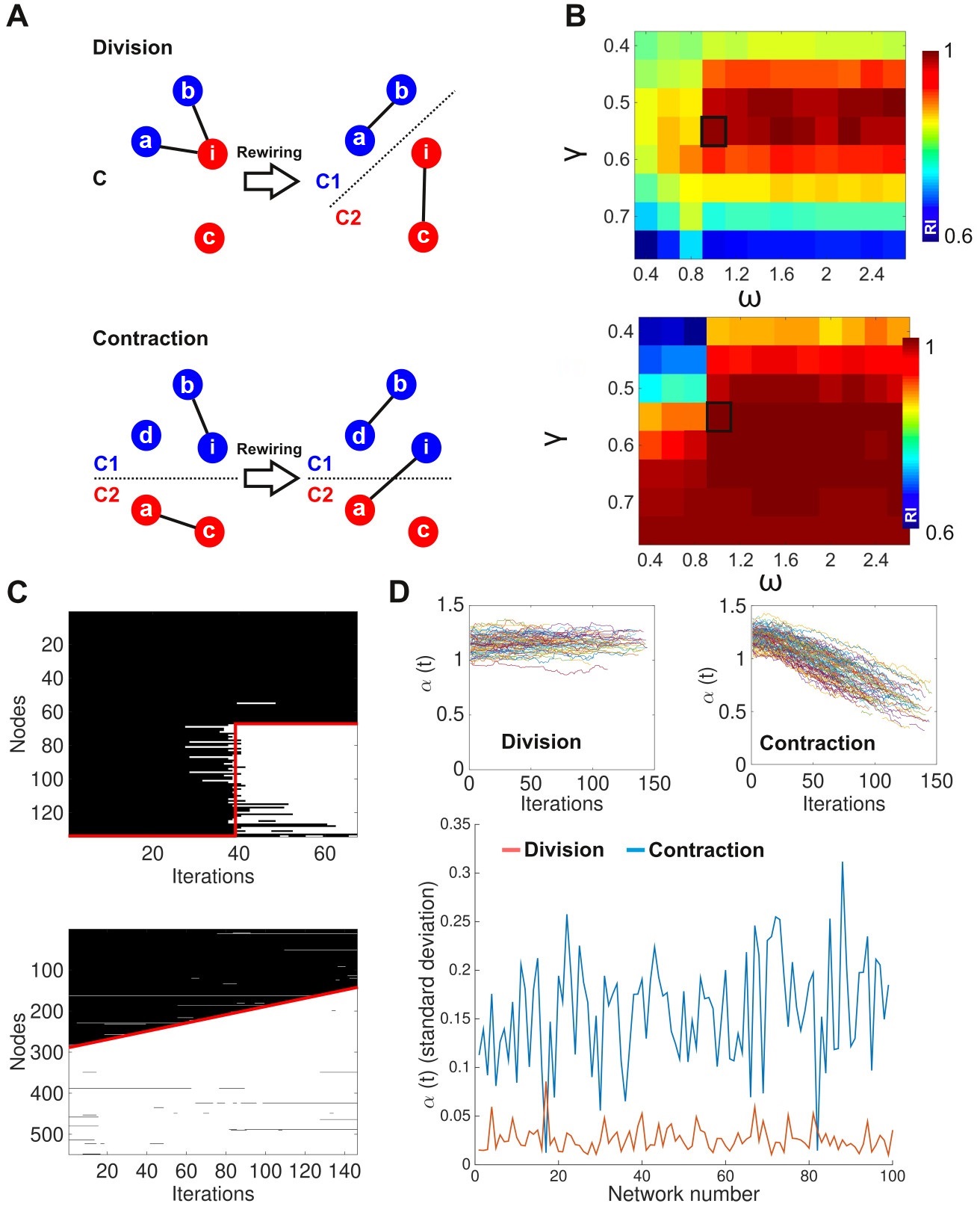}

     \caption{Benchmark for time-dependent heterogeneous networks based constructed from two dynamic processes (split and merge). (A) Rewiring steps used to generate the dynamics of division and contraction of communities. (B) Rand index values between the modules detected by the multilayer  Louvain algorithm and the theoretical modules averaged over 500 networks, for different values of $\gamma$ and $\omega$. Optimal performance was found for $\gamma=0.55$ and $\omega=1$. (C) Modular structure detected by the multilayer Louvain algorithm using the optimal parameters for merge and split processes. The red lines indicate the expected distribution of module membership labels. (D) The coefficient of the power law for the degree distribution, $\alpha(t)$, and the standard deviation of $\alpha$ for both dynamics vs. time.} % Figure caption

	\label{fig3}
\end{figure}

We present an algorithm to induce this type of dynamics in heterogeneous networks of $N$ nodes constructed according to the algorithm presented by Lancichinetti et al.\cite{lancichinetti2008benchmark}. We start from a network with $N$ nodes and apply the following steps:

\begin{enumerate}
%\item A module larger than $N \times p$ (C) is chosen at random, where $p$ (the fraction of nodes in the network) is such that $N \times p$ is sufficiently large for the division into smaller sub-modules.
\item A module named C, sufficiently large for the division into smaller sub-modules, is chosen at random.

\item The nodes belonging to C are assigned to two sub-modules, C1 and C2. A fraction $x$ of the nodes in C belongs to C1, and $(1-x)$ belongs to C2.

\item For each node $i$ in C2, we apply the following steps:

\begin{enumerate}

	\item $\mu_{21i}$ of node i is calculated as:
	
	 $\mu_{21i} = \frac{\# \text{Links with C1 nodes}}{\# \text{Total links}}  $
	 
    \item A mixing parameter between communities is chosen per node, $\mu_{12}$.
    
    \item While $\mu_{21i}>\mu_{12}$ nodes a, b and c are searched such that they meet the conditions set forth in the division step presented in Fig \ref{fig3}A (division) : node a and node b belong to C1 and are connected to node i, node c belongs to C2 and is not connected to node i.
    
    \item The link between node i and node b is deleted and a new link is created connecting it to node a. This is repeated until there are no more nodes a, b and c meeting these conditions or until $ \mu_{21i} < \mu_{12} $. The rewiring scheme is presented in Fig. \ref{fig3}A.
    
    \item The final adjacency sub-matrix is saved as the dynamic network at time $t=i$.
    
\end{enumerate}
\end{enumerate}

\subsubsection{Null model based on contraction of communities: grow-shrink}\label{grow_shrink}

We also present an algorithm that selects two random modules from a certain complex network with N nodes, and then grows the size of one module at the expense of the other. The steps are as follows:

\begin{enumerate}
    
%\item Two modules C1 and C2 of sizes larger than N*p are chosen at random. P is taken as a fraction such that N*p is large enough to allow the temporal evolution of the network with a significant number of nodes involved.

\item Two modules C1 and C2 are chosen at random.

\item The adjacency matrix is reordered so that the two modules appear consecutively. The adjacency submatrix containing modules C1 and C2 is selected.

\item C1 nodes are re-tagged and rewired, changing their membership to C2: Let $ n_1$ be the number of C1 nodes, ranging from 1 to the $X-th$ node of C1, where  X is the integer part of $n_1x$. The following steps are applied to each node $i$ of this set: 

\begin{enumerate}
    
	\item $ \mu_{12i}$ of node i is calculated as,
	
	 $  \mu_{12i} = \frac{\# \text{Links with C2 nodes}}{\# \text{Total links}}  $
	
		\item A mixing parameter between modules is chosen per node, $ \mu_{12} $
		
		\item While $ \mu_{12i} > \mu_{12} $, nodes a, b, c and d are searched 
		such that they meet the conditions set forth in Fig \ref{fig3}A (contraction).
		
		\item The link from node i to node b is deleted, and a new link is created between node i and node a. This is repeated until there are no more nodes a and b fulfilling these conditions, or until $ \mu_{12i} < \mu_{12} $. The rewiring scheme is presented in Fig. \ref{fig3}A.
		
\item  Node i is removed from C1 and added to C2.
\item  The average degree of intercommunity links of C2 nodes ($ <k_{iC2}>$) and the degree of node i ($k_i$) are calculated.

\item  While $k_i <$ $<k_{iC2}>$ pairs of nodes whose C2 intramodular degrees are between $<k_{iC2}>$ and $k_{max}$ are selected. The links between those nodes are deleted, and new links between node i and other nodes in C2 with $ k <$ $<k_{iC2}>$ are added.

\item The final adjacency submatrix is saved as the dynamic network at time $t=i$.
\end{enumerate}
\end{enumerate}

\subsection*{Construction of dynamic networks from fMRI data}

After preprocessing, we extracted the average BOLD signals from 401 regions of interest, for all conditions and participants. These regions correspond to a sub-parcellation of the automated anatomic labeling (AAL) atlas\citep{Tzourio-Mazoyer2002}, resulting in more detailed anatomical subdivisions\cite{crossley2014hubs}. Next, we applied a sliding window procedure to compute Pearson's linear correlation coefficient between all pairs of ROIs during 60 s windows, moving each window for one volume to compute the next correlation value. This resulted in a weighted correlation matrix $C_{ijs}$ containing in its $i,j,s$ the correlation coefficient between ROIs $i$ and $j$ during a 60 s window starting at volume $s$. Since the multilayer Louvain algorithm requires positive link weights, we opted to threshold the correlation matrix to produce a binary adjacency matrix $A_{ijs}$ which equals 1 only if correlation between nodes $i,j$ at volume $s$ is larger than a certain threshold, and zero otherwise. Since we used our dynamic benchmark to estimate to optimal $\gamma$ and $\omega$ values, we chose a threshold to result in the same average degree used for the benchmark ($<k>=20$), so that both networks were comparable. 

\subsection*{Node flexibility and the largest multilayer module}

We defined two metrics based on the module membership matrix $G_{it}$ given by the multilayer Louvain algorithm. First, we defined the node flexibility $F_i$ as the normalized number of module membership changes for that node:

\begin{equation}
F_i = \frac{|\{t : G_{i t} \neq G_{i t+1} \}|}{T}
\end{equation}

where $T$ is the total number of volumes in the recording.

We defined the largest multilayer module size $LMM$ as the normalized size of the largest module in $G_{it}$, i.e. 

\begin{equation}
LMM = \frac{\max_{i} |G_{i t}|}{NT}
\end{equation}

where $N$ is the number of nodes and $T$ the total number of volumes in the recording, so $NT$ is the maximum possible size for the giant module.

\section*{Results}

\subsection*{Time-dependent benchmark and parameter selection}

We investigated the performance of the multilayer Louvain algorithm based on introducing equally weighted ($\omega$) connections between consecutive temporal layers (see Fig. \ref{figure1}B). Also, we considered equal resolution parameters across layers ($\gamma$). Thus, the module detection algorithm depended only on these two parameters.

We applied our benchmark to determine the optimal parameter values. For this purpose, we introduced a grid of $\gamma$ and $\omega$ values, and for each pair of values we measured the Rand index between the ground-truth modules and those detected using the multilayer Louvain algorithm, averaging the results over 500 independent realizations (see Fig. \ref{fig3}B). The ground truth used to determine the performance of the multilayer  Louvain algorithm was determined by  applying the standard Louvain algorithm to each time point and then selecting the partition maximizing the overlap with the heuristic expectation. It is important to note that this could only be done for networks with \emph{a priori} information about the allocation of nodes in temporal modules, such as in the merge-split and grow-shrink dynamics.

The optimal parameters obtained following this procedure were $\gamma=0.55$ and $\omega=1$ for both benchmarks (values are indicated as black boxes in Fig. \ref{fig3}B). 

Finally, while the merge-split and grow-shrink dynamics preserved the amount of total links in the network, the degree distributions could be changed during temporal evolution. To quantify this possibility we measured  $\alpha(t)$, i.e. the time-dependent exponent of the power law degree distribution, and then computed its standard deviation across time for both dynamics (see Fig. \ref{fig3}D). We observe that $\alpha (t)$ changed for the contraction dynamics, while it remained stable for the division dynamics.

\subsection*{Modular structure of dynamic brain connectivity networks}

We applied the multilayer Louvain algorithm using the optimal parameters inferred from the benchmarks to dynamic functional connectivity networks obtained from fMRI data. First, we investigated healthy subjects during conscious wakefulness, and during the wake-sleep cycle, consisting of four stages of increasing depth and progressive loss of conscious content (N1, N2 and N3, respectively).  

\begin{figure}[t]
	\centering
	\includegraphics[width=\linewidth]{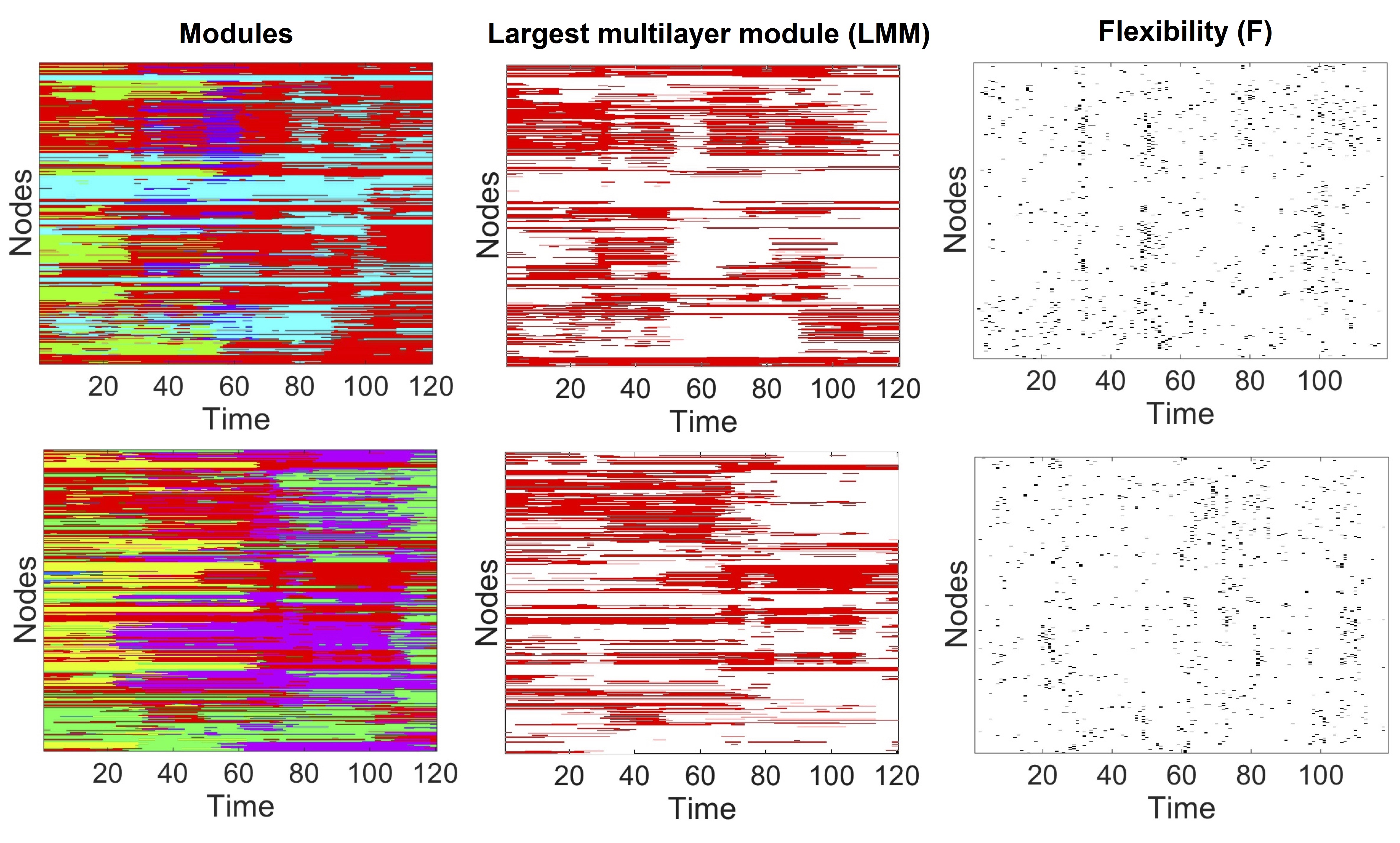}

 	\caption{Temporal modules computed from a single subject during wakefulness (first row) and N3 sleep (second row), detected using the multilayer Louvain algorithm with $ \gamma = 0.55 $ and $ \omega = 1 $. Colors represent different modules.  The largest multilayer module (LMM) is shown in red color. Right panels represent the transition points between modules, which are used to compute the flexibility (F) of each node.}

	\label{fig4}
\end{figure}

%For fMRI dynamic networks to be comparable with our benchmark, we thresholded them to guarantee the same average degree $<k>$ = 20.  ESTO LO DIJIMOS ANTES
Fig. \ref{fig4} shows sample temporal modules detected from a single subject during wakefulness and N3 (first and second rows, respectively), as well as the largest multilayer module (LMM) for each condition. The right panels indicate the transition points between modules, which are used to compute the flexibility (F) of each node.

\begin{figure}[t]
	\centering
	\includegraphics[width=\linewidth]{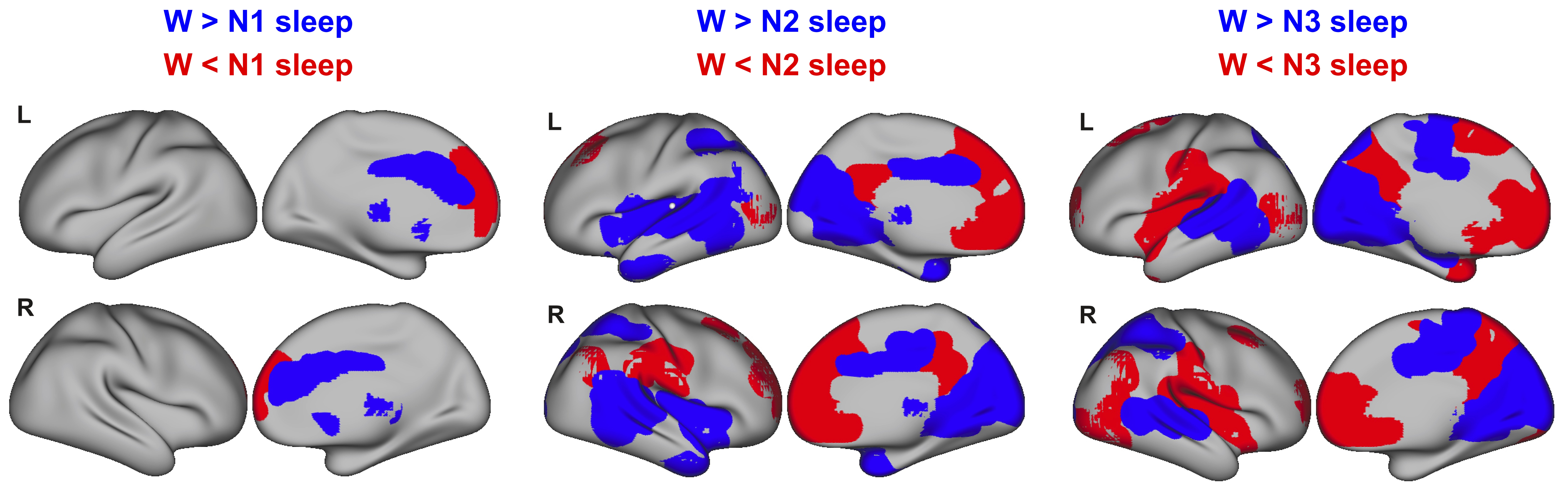}

 	\caption{Comparison of node flexibilities between wakefulness and sleep (N1, N2 and N3 sleep). Statistical significance was determined by Student's t-test (p$<$0.05, FDR corrected for multiple comparisons).}

	\label{fig5}
\end{figure}

We then compared the node flexibility between wakefulness and each sleep stage (Fig. \ref{fig5}). Brain areas corresponding to nodes with decreased flexibility vs. wakefulness are shown in blue, while those with increased flexibility vs. wakefulness are shown in red (p$<$0.05, FDR corrected). Results in Fig. \ref{fig5} show that regions presenting decreased flexibility during sleep are related to sensory perception, and also included subcortical regions that serve as intermediate stages for the propagation of sensory information towards the cortex, such as the thalamus. In particular, flexibility decreases during N1 sleep were observed mostly for thalamic and anterior cingulate nodes, consistent with the observation that the thalamus becomes deactivated and disconnected from sensory cortices during early sleep\cite{tagliazucchi2014decoding}. Conversely, flexibility increased during sleep in parietal, frontal and prefrontal regions associated with higher cognitive functions, with the strongest increases seen during N3 sleep. 

\begin{figure}[t]
	\centering
	\includegraphics[width=12cm]{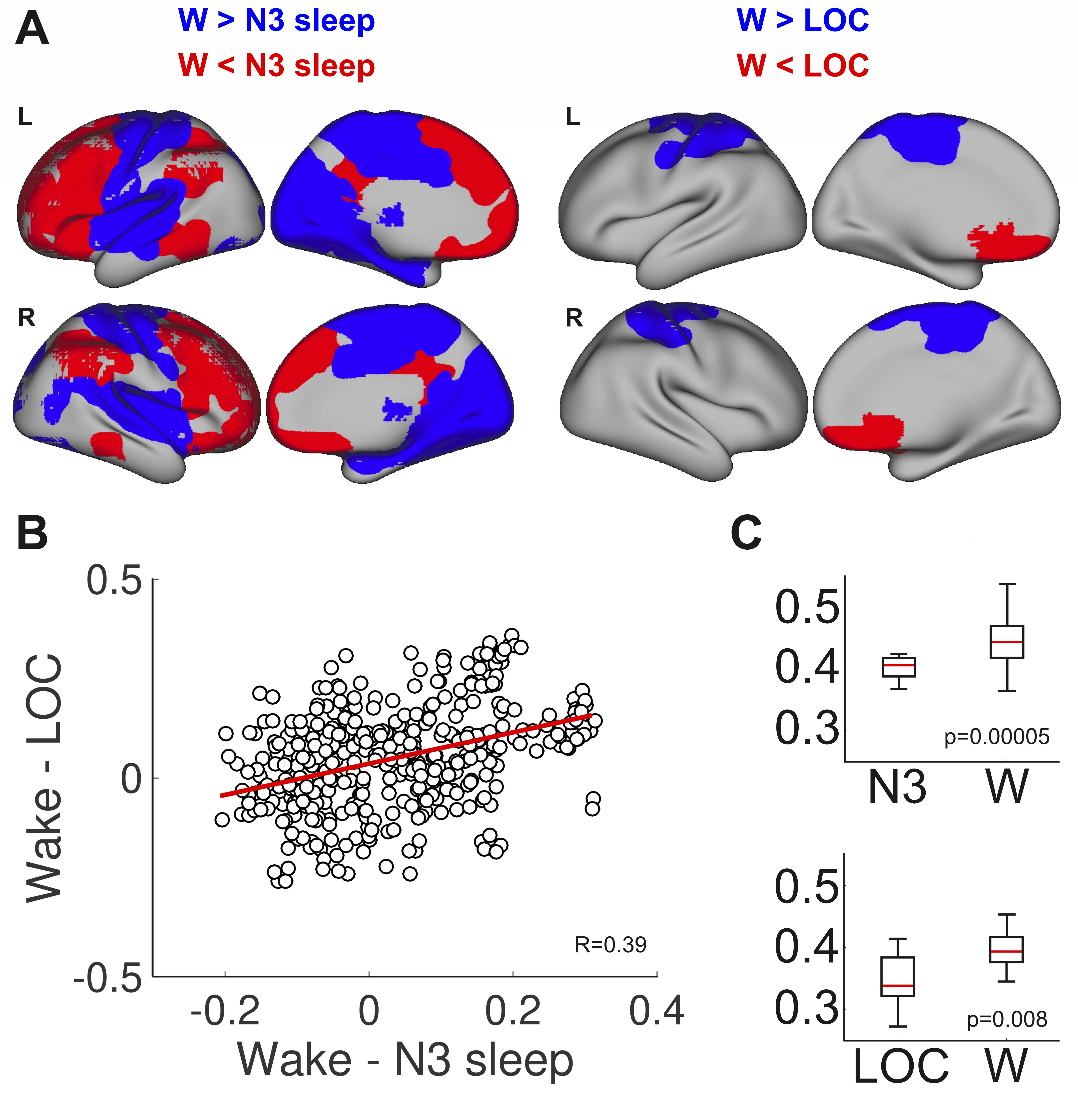}

 	\caption{(A) Significant differences in the regional probability of belonging to the largest multilayer module for wakefulness vs. N3 sleep (left) and vs. LOC (right). (B) Scatter plot of changes in this probability for N3 vs. LOC, where each point represents a node in the parcellation. Even though less nodes presented significant changes for LOC compared to N3, the spatial pattern of the changes was similar, as determined by the correlation coefficient (R=0.39). (C) Boxplots for the normalized size of the largest multilayer module, both for N3 vs wakefulness (top) and N3 vs. LOC (bottom). In both cases, the largest module decreased in size during loss of consciousness.}

	\label{fig6}
\end{figure}

Finally, we compared the regional probability of belonging to the largest multilayer module in wakefulness vs. deep sleep and propofol-induced loss of consciousness (LOC). Figure \ref{fig6}A presents a comparison of significant changes in this probability between wakefulness, N3 and LOC. While changes were more widespread and significant during N3 sleep, LOC was also associated with decreases in sensorimotor regions, and increases in frontal regions. A scatter plot of the change in the probability of belonging to the largest multilayer module for LOC vs. N3 shows that even though less regions were significant for LOC, the pattern of changes was similar to that measured during N3 sleep (Fig. \ref{fig6}B). Also, both N3 and LOC were characterized by smaller sizes of their largest multilayer modules. As shown in Fig. \ref{fig6}C, the normalized size of the largest modules was significantly smaller for N3 and LOC compared to wakefulness.

\section*{Discussion}

The human brain is inherently dynamic, constantly presenting non-trivial scale-free fluctuations both in space and time \cite{cavanna2018dynamic,he2014scale}. During past years, several studies have demonstrated the usefulness of adopting a dynamic approach towards brain activity \cite{hutchison2013dynamic}, shedding light on the large-scale brain dynamics of physiological and pathological brain states \cite{gifford2020resting}, the coordination between distributed networks of regions that gives rise to human cognition and behavior \cite{braun2015dynamic, telesford2016detection, bassett2011dynamic, pedersen2018multilayer}, and the relationship between multimodal brain activity \cite{tagliazucchi2012dynamic}, anatomy \cite{liegeois2016cerebral}, and genetics \cite{diez2018neurogenetic}. Global states of consciousness such  sleep, general anesthesia, and disorders of consciousness have also been explored from this perspective, leading to the general observation of a reduced and more stable repertoire of brain state associated with unconsciousness \cite{barttfeld2015signature, tagliazucchi2017large,huang2020temporal, demertzi2019human,ma2017dynamic,cavanna2018dynamic}.  

These results depend on the development of new methods capable of extracting temporally recurrent brain activity patterns in data sets comprising hundreds of thousands of voxels. At the core of dynamic analyses of fMRI data is the need for data-driven identification of recurrent brain states, and their mutual transition probabilities. This can be achieved by different means, such as clustering a point process representation of the data obtained by thresholding the BOLD signals \cite{tagliazucchi2012criticality,tagliazucchi2016voxel, liu2018co}, or applying more sophisticated methods based on hidden Markov chains \cite{hunyadi2019dynamic}. One of the first and most frequently applied methods to unveil the dynamics of whole-brain fMRI activity consists of adopting a multi-layer network representation, where each layer corresponds to a different time window and intralayer connections indicate BOLD signal correlation (i.e. functional connectivity) \cite{muldoon2016network}. By introducing interlayer connectivity, modularity optimization algorithms can then be applied to this representation for the data-driven discovery of temporal modules, i.e. non-overlapping groups of anatomical regions presenting high and low intra- and inter-group functional connectivity, respectively \cite{muldoon2016network,betzel2017multi}. 

Module detection in temporal networks remains one of the most used methods to map the dynamic evolution of whole-brain activity, yet presents shortcomings related to the need to adjust free parameters. Our contribution advances our understanding of whole-brain dynamics from the two aforementioned perspectives. First, we developed a systematic framework to evaluate the performance of modularity optimization algorithms by constructing a dynamic benchmark with heterogeneous and realistic node distributions, allowing the justified choice of model parameters. Next, we started from these parameters to show that sleep progressively stabilizes module assignments in sensory  and associative cortical regions, with the opposite behavior observed for frontal, parietal and temporal regions spanning key default mode network hubs. Sleep and propofol anesthesia reduced the regional probability of belonging to the largest multilayer module for most occipital, temporal and parietal brain regions, with these changes being correlated between two unconscious conditions, suggesting an overlapping mechanism implicated in loss of consciousness. 

The size of the largest multilayer module represents a new metric to quantify integration simultaneously in the spatial and temporal domains. Since modules are computed from the multilayer representation of the network, this metric is different from the largest module at any given time point. The largest module can shrink to a few nodes or regions at a certain time, but grow at later times to encompass most of the brain; in other words, the \emph{identity} of the module over time is taken into account. This is in line with the definition of the dynamic core by Edelman and Tononi, which corresponds to the \emph{same temporal process} which involves different anatomical regions at different times, but nevertheless maintains its identity through time \cite{edelman2000reentry,edelman2008universe,tononi1998consciousness}. Accordingly, consciousness can vanish as consequence of a static or fragmented dynamic core (see Figure \ref{figure1}). The reduced size of the largest multilayer module during deep sleep and general anesthesia module seemingly supports the second possibility; furthermore, the regional correlation between the changes in the likelihood of belonging to the largest module during deep sleep and anesthesia suggests that this metric could capture a signature of fading consciousness present in both conditions.

In terms of flexibility, i.e. switching rate between module assignment, we did not observe significant changes at the global level, but detected two sets of regions with opposite changes during the descent to deep sleep. This dissociation became especially evident during N3 sleep, when flexibility increased in frontal and parietal regions, and decreased in primary sensory regions; these patterns persisted during N3 or slow wave sleep. We can speculate that the progressive fragmentation of largest module results in increased switching behavior in fronto-parietal regions, while the stabilization of module dynamics in sensory regions could be the results of neurophysiological processes preventing arousals from sleep, such as K-complexes and slow waves \cite{caporro2012functional,jahnke2012wake,dang2008spontaneous}. These processes seem specific to sleep, since general anesthesia did not results in local flexibility changes. 

While this is the first study applying a multilayer network approach to understand the temporal dynamics of fMRI data during unconsciousness, there is one antecedent following similar methods applied to EEG data measured in patients suffering from disorders of consciousness \cite{cai2020characterization}. This work demonstrated decreased flexibility in relatively slow frequency bands (mainly in the alpha range) and the opposite behavior over longer time scales. Our work might represent the counterpart of these findings in the hemodynamic activity measured using fMRI during sleep \cite{mantini2007electrophysiological}, where certain networks maintain high levels of network flexibility to revert towards a state of wakefulness upon arousal (as opposed to brain-injured patients who might not spontaneously revert towards conscious wakefulness).

Future research using multilayer network modularity to describe temporal fMRI dynamics should base parameter selection on an adequate benchmark. Using model networks with known behavior is a widely adopted practice in many applications of network science; ideally, this model should represent as many features of the empirical networks as possible \cite{lancichinetti2008benchmark}. Our methodological development filled this gap by introducing a benchmark for temporal modular networks with heterogeneous connectivity giving rise to realistic node and module size distributions. This benchmark can also find future applications in the comparison of different methods for modularity detection, and in the construction of synthetic datasets with known modular structure to serve as null models.

In summary, we investigated for the first time modular brain network dynamics during states of unconsciousness, finding converging evidence of a reconfiguration of the largest multilayer module during deep sleep and general anesthesia. We interpreted these changes in the light of the dynamic core theory, concluding that unconsciousness results in its fragmentation in spite of preserved stability. Future studies should assess whole-brain dynamics simultaneously with different methods to understand whether the dynamic core fluctuates over scales inaccessible to fMRI, and whether these fluctuations are manifest at the behavioral and cognitive levels.   

\section*{Acknowledgements}

This  work  was  supported  by  funding  from  Agencia  Nacional De Promocion Cientifica Y Tecnologica (Argentina), grant PICT-2018-03103.

%%%%%%%%%%%%%%%%%%%%%%%%%%%%%%%%%%%%%%%%%%%%%%%%%%%%%%%%%%%%
%%%%%%%%%%%%%%%%%%%%%%%%%%%%%%%%%%%%%%%%%%%%%%%%%%%%%%%%%%%%
\bibliography{bibliography}

%%%%%%%%%%%%%%%%%%%%%%%%%%%%%%%%%%%%%%%%%%%%%%%%%%%%%%%%%%%%
%%%%%%%%%%%%%%%%%%%%%%%%%%%%%%%%%%%%%%%%%%%%%%%%%%%%%%%%%%%%

%%%%%%%%%%%%%%%%%%%%%%%%%%%%%%%%%%%%%%%%%%%%%%%%%%%%%%%%%%%%
%%%%%%%%%%%%%%%%%%%%%%%%%%%%%%%%%%%%%%%%%%%%%%%%%%%%%%%%%%%%
%\section*{Author contributions statement}
\end{document}

\renewcommand{\figurename}{Supplementary Figure}
\renewcommand{\tablename}{Supplementary Table}
\flushbottom
\maketitle
\thispagestyle{empty}
\newpage

\subsection*{Supplementary Subsection 1. Heterogeneous entropy changes induced by 5HT2A-R activation on anatomical and functional groupings of brain regions.}
% Figura del sgrupamiento anatomico
We asked whether the heterogeneous changes of entropy induced by 5HT2A-R activation could be explained by grouping the AAL brain regions according to anatomical and functional criteria. 
Regarding anatomical criteria brain regions can be spatially split into 8 major non-overlapping anatomical groups: Frontal, Temporal, Parietal, Occipital, Limbic, Sensorimotor, Cingulate and Subcortical.\cite{Tzourio-Mazoyer2002} 
Regarding the functional grouping, we used the Resting State Networks, a functional grouping of brain regions based on the observed spatio-temporal patterns of BOLD signals during resting state activity.\cite{Beckmann2005} 
These groups are Salience (Sal), Fronto-Parietal (FPN), Default Mode (DMN), Primary Visual (Vis), Extrastiate Cortex (EC), Auditory (Aud), Sensorimotor (SM),  and Executive Control (EC). The first one, Sal, was obtained from Lee\textit{et al}.~\cite{Lee2017}, the second and third, FPN and DMN, were obtained from Oliver~\textit{et al}.\cite{Oliver2019}, and the rest from Beckmann \textit{et al}.~\cite{Beckmann2005} 
Brain regions can potentially belong to different functional groups.
%In Fig.Supplementary Figure 1 we present the $\Delta H$ value for the regions belonging to each of the anatomical groups (Supplementary Figure 1), showing that 

The effect of 5HT2A-R activation on entropy is heterogeneous also at the level of anatomical groups (Supplementary Figure~\subref{fig:anatfunc}{A}) -- i.e. within all the groups there were both regions with increased and decreased entropy after the 5HT2A-R activation. However, Occipital and Cingulate regions show a strong tendency to increase their entropy, in agreement with entropy increases in these regions observed in human experiments with serotonergic psychedelics.\cite{Schartner2017,Lebedev2016}
Regarding the functional grouping (Supplementary Figure~\subref{fig:anatfunc}{B}), we also found that the effect of 5HT2A-R activation on regional entropy (Fig. 3B) is heterogeneous within groups, with the exception of Vis and FPN, where almost all the regions increased and decreased their entropy, respectively.
Note also that with the exception of the angular gyri, all the DMN regions increase their entropy, which resonates with the observed reduction of the DMN integrity on humans during psychedelic experiences.\cite{Carhart-Harris2016}

\begin{figure}[th!]
	\centering
	\includegraphics[scale=0.5]{./figures/supp_fig1.png}

  \caption{\textbf{The effect of 5HT2A-R activation is not explained by anatomical or functional groupings.}
	(\textbf{A}) 5HT2A-R activation heterogeneously change entropy in regions belonging to the same anatomical group.
	(\textbf{B}) Entropy also changed in a heterogeneous way when functional grouping is considered. Note the special case of FPN and Vis networks, for which almost all regions decreased and increased their entropy, respectively (note that functional groups can be overlapping).
	Circles are averages and error bars 1 s.d. computed from 1000 simulations.}

	\label{fig:anatfunc}
\end{figure}

\newpage
\subsection*{Supplementary Subsection 2. Connectivity strength is the best predictor for entropy changes among local connectivity measures.}

We control the role of local connectivity on the observed entropy changes ($\Delta h_n$) induced by 5HT2A-R activation using as predictors for $\Delta h_n$ other local connectivity measures as: degree,\cite{Zuo2012} eigenvector centrality,\cite{Zuo2012} communicability,\cite{Andreotti2014} page-rank,\cite{Zuo2012} sub-graph centrality,\cite{Zuo2012} and closeness centrality.\cite{Rubinov2010}
We confirmed that local connectivity strength is the best linear predictor of entropy changes among other centrality measures.

\begin{figure}[th!]
	\centering
	\includegraphics[width=\linewidth,trim={3cm 0 3cm 0},clip]{./figures/supp_fig2.png}

  \caption{\textbf{Centrality measures as possible explanatory variables for $\Delta h_n$}.
  Local connectivity strength is the best linear predictor of $\Delta h_n$ among other centrality measures. Communicability and sub-graph centrality also show a good linear relationship with $\Delta h_n$, though with higher residual variance. Circles are averages and error bars 1 s.d. out of 1000 independent simulations.}

	\label{fig:centrality}
\end{figure}

\newpage
\subsection*{Supplementary Subsection 3. Receptors controls.}

We control the role of receptor density distribution on the observed  entropy changes ($\Delta h_n$) induced by 5HT2A-R activation by using alternative receptor distributions. 
To test the effect of the heterogeneous 5HT2A-R distribution on $\Delta h_n$, we used an uniform receptor distribution, where the receptor density for each region corresponds to the average 5HT2A-R density among all regions. 
To evaluate the role of the specific 5HT2A-R distribution, a randomised version of the 5HT2A-R was also used, where for each simulation the receptor expression is randomised.
Finally, to evaluate the effect of activating other serotonin receptors, we used the 5HT4-R distribution, a receptor thought to play no role on the psychedelic state. 

On the one hand, the Uniform and Random receptor distributions show almost the same behaviour on average, with $\Delta h_n$ showing a linear relationship with the node strength that changed slope at node strength value $\sim 0.35$. 
The Random distribution shows more variability because the receptor distribution is randomised on each simulation.
On the other hand, the 5HT4-R distribution changes entropy but in a subtler manner, yielding an average $\Delta h_n \sim 0$.

On summary, the heterogeneous 5HT2A-R distribution critically impacts the heterogeneity of $\Delta h_n$ values and its relationship with the local connectivity strength. In addition, using other empirically-based distribution yields no significant change in average brain entropy, highlighting the key relationship between 5HT2A-R density and connectivity on the entropic effects induced by serotonergic psychedelic drugs.

% We confirmed that local connectivity strength is the best linear predictor of entropy changes among other centrality measures.

\begin{figure}[th!]
	\centering
	\includegraphics[width=\linewidth,trim={3cm 5cm 3cm 5cm},clip]{./figures/supp_fig5.png}

  \caption{\textbf{Controls for receptor distribution vs $\Delta h_n$}.
  $\Delta h_n$ vs connectivity strength is plotted for 5HT2A-R, Uniform, 5HT4-R, and Random distribution of receptors (see text for details).
  The specific distribution of 5HT2A-R is critical both for the heterogeneity of $\Delta h_n$ and for the average increase in local entropy.
    Circles are averages and error bars 1 s.d. out of 1000 independent simulations.
  }

	\label{fig:receptores}
\end{figure}

\newpage
\subsection*{Supplementary Subsection 4. The probability distribution of excitatory firing rates can be well fitted by a Gamma distribution.}

To check the goodness of fit (GOF) of Gamma distribution to the simulated excitatory firing rates of each region, we generated $10^6$ simulation points for each region under both PLA and 5HT2A condition, then, a Gamma distribution was fitted and the Kolmogorov-Smirnov (K-S) distance between the firing rate distribution and 1000 random samples (same size) of the respective fitted Gamma distribution was computed.
We summarise the Gamma GOF as the average K-S for each region under both conditions.
To assess the significance of average GOF values, we generated an acceptance interval computing the K-S distance between 100 independent samples (same size) sampled from exactly the same Gamma distribution (re-sampled K-S). 
This procedure was repeated for different set of Gamma distribution parameters.
If the average K-S distance of a given region falls below the maximum re-sampled K-S distance, we consider this region to be well fitted by the Gamma distribution.

Despite the decreased GOF under 5HT2A condition (compared to PLA) all regions fall within the acceptance interval under both conditions, enabling us to use the Gamma distribution parameters to estimate each region's Shannon's differential entropy.

\begin{figure}[th!]
	\centering
	\includegraphics[width=\linewidth,trim={3cm 0 3cm 0},clip]{./figures/supp_fig3.png}

  \caption{\textbf{Goodness of fit of Gamma distribution to firing rates distribution are within confidence intervals.}
  Distribution of K-S distance values for PLA (top) and 5HT2A-R (bottom) condition for each region are represented by violins, where white circle denotes average K-S. Black (red) dashed line represent the average (maximum) re-sampled K-S distance (see the text for details). }

	\label{fig:gammadist}
\end{figure}

\newpage
\subsection*{Supplementary Subsection 5. Brain Regions of the Automated Anatomical Labeling.}
\begin{table}[th!]
\centering
\caption{AAL regions and abreviations}
\label{tab:abreviations}
    \begin{tabular}{ c | c }
    % \hline
    AAL Abbreviation & Brain Regions \\
    \toprule
    % \hline \hline
    \hline
    Precentral & Precental gyrus \\
    \hline
    Front Sup & Superior frontal gyrus, dorsolateral\\
    \hline
    Front Sup Orb & Superior frontal gyrus, orbital part\\
    \hline
    Front Mid & Middle frontal gyrus \\
    \hline
    Front Mid Orb & Middle frontal gyrus, orbital part\\
    \hline
    Front Inf Oper & Inferior frontal gyrus, opercular part\\
    \hline
    Front Inf Tri & Inferior frontal gyrus, triangular part\\
    \hline
    Front Inf Orb & Inferior frontal gyrus, orbital part\\
    \hline
    Rolandic Oper & Rolandic operculum \\
    \hline
    Supp Motor Ar & Supplementary motor area \\
    \hline
    Olfactory & Olfactory cortex \\
    \hline
    Frontal Sup Med & Superior frontal gyrus, medial \\
    \hline
    Frontal Mid Orb & Superior frontal gyrus, medial orbital\\
    \hline
    Rectus & Gyrus rectus \\
    \hline
    Insula & Insula \\
    \hline
    Cingulum Ant & Anterior cingulate and paracingulate gyri\\
    \hline
    Cingulum Mid & Median cingulate and paracingulate gyri\\
    \hline
    Cingulum Post & Posterior cingulate gyrus \\
    \hline
    Hippocampus & Hippocampus \\
    \hline
    ParaHippocamp & Parahippocampal gyrus \\
    \hline
    Amygdala & Amygdala \\
    \hline
    Calcarine & Calcarine fissure and surrounding cortex\\
    \hline
    Cuneus & Cuneus \\
    \hline
    Lingual & Lingual gyrus \\
    \hline
    Occipital Sup & Superior occipital gyrus \\
    \hline
    Occipital Mid & Middle occipital gyrus \\
    \hline
    Occipital Inf & Inferior occipital gyrus \\
    \hline
    Fusiform & Fusiform gyrus \\
    \hline
    Postcentral & Postcentral gyrus \\
    \hline
    Parietal Sup & Superior parietal gyrus \\
    \hline
    Parietal Inf & Inferior parietal, but supramarginal and angular gyri\\
    \hline
    SupraMarginal & Supramarginal gyrus \\
    \hline
    Angular & Angular gyrus \\
    \hline
    Precuneus & Precuneus \\
    \hline
    Paracentral Lobule & Paracentral lobule \\
    \hline
    Caudate & Caudate nucleus \\
    \hline
    Putamen & Lenticular nucleus, putamen \\
    \hline
    Pallidum & Lenticular nucleus, pallidum \\
    \hline
    Thalamus & Thalamus \\
    \hline
    Heschl & Heschl gyrus \\
    \hline
    Temporal Sup & Superior temporal gyrus \\
    \hline
    Temporal Pol Sup & Temporal pole: superior temporal gyrus\\
    \hline
    Temporal Mid & Middle temporal gyrus \\
    \hline
    Temporal Pol Mid & Temporal pole: middle temporal gyrus\\
    \hline
    Temporal Inf & Inferior temporal gyrus \\
    \hline
    
    \end{tabular}
\end{table}

\newpage
\subsection*{Supplementary Subsection 6. Anatomical and Functional Grouping of Brain Regions.}

\begin{table}[ht!]
\centering
\caption{Anatomical Grouping}
\label{tab:anatomical}
    \begin{tabular}{c|c}
    Anatomical Group & AAL Regions \\
    \toprule
    \multirow{3}{*}{Cingulate} & Cingulum Ant\\
    & Cingulum Mid \\
    & Cingulum Post \\
    \hline
    
    \multirow{12}{*}{Frontal} & Frontal Sup\\
    &Front Sup Orb\\
    &Front Mid    \\
    &Front Mid Orb\\
    &Front Inf Ope\\
    &Front Inf Tri\\
    &Front Inf Orb\\
    &Supp Motor Ar\\
    &Olfactory    \\
    &Front Sup Med\\
    &Front Med Orb\\
    &Rectus\\
    \hline
    
    \multirow{3}{*}{Limbic} & Hippocampus\\
    &ParaHippocamp\\
    &Amygdala     \\
    \hline
    
    \multirow{7}{*}{Occipital} & Calcarine\\
    &Cuneus       \\
    &Lingual      \\
    &Occipital Sup\\
    &Occipital Mid\\
    &Occipital Inf\\
    &Fusiform     \\
    \hline
    
    \multirow{6}{*}{Parietal} & Parietal Sup\\
    &Parietal Inf \\
    &SupraMarginal\\
    &Angular      \\
    &Precuneus    \\
    &Paracentr Lob\\
    \hline
    
    \multirow{7}{*}{Sensorimotor} & Parietal Sup\\
    &Parietal Inf \\
    &SupraMarginal\\
    &Angular      \\
    &Precuneus    \\
    &Paracentr Lob\\
    \hline
    
    \multirow{4}{*}{Subcortical} & L Caudate \\
    &Putamen\\ 
    &Pallidum\\
    &Thalamus\\
    \hline
    
    \multirow{7}{*}{Temporal} & Insula\\
    &Heschl       \\
    &Temporal Sup \\
    &Tempr Pol Sup\\
    &Temporal Mid \\
    &Tempr Pol Mid\\
    &Temporal Inf \\
    \hline

    \end{tabular}
\end{table}
\\
\begin{table}[ht!]

    \centering
    \caption{Functional Grouping. If no hemisphere is specified, then regions from both left (L) and right (R) hemisphere are used.}
    \label{tab:functional}
    \begin{adjustbox}{scale=0.8,center}
        \begin{tabular}{c|c}
        Functional Group & AAL Regions \\
        \toprule
        \multirow{4}{*}{Visual (Vis)} & Calcarine\\
        & Cuneus \\
        & Lingual \\
        & Occipital Sup (L) \\
        \hline
        
        \multirow{5}{*}{Extrastriate (ES)} & Lingual (L)\\
        &Occipital Sup\\
        &Occipital Mid\\
        &Occipital Inf\\
        &Parietal Sup\\
        \hline
        
        \multirow{13}{*}{Auditory (Aud)} &  Front Inf Ope (L) \\
        &Front Inf Tri (L)\\
        &Rolandic Oper\\
        &Insula       \\
        &Cingulum Ant \\
        &Cingulum Mid \\
        &Amygdala     \\
        &SupraMarginal\\
        &Putamen      \\
        &Pallidum     \\
        &Thalamus     \\
        &Heschl       \\
        &Temporal Sup \\
        &Temporal Mid (R)\\
        &Tempr Pol Sup (R)\\
        \hline
        
        \multirow{8}{*}{Somatomotor (SM)} & Heschl\\
        &Paracentr Lob\\
        &Parietal Sup \\
        &Postcentral  \\
        &Cingulum Mid \\
        &Rolandic Oper\\
        &Precentral (R)\\   
    
        \hline
        
        \multirow{13}{*}{Excecutive Control (EC)} & Precentral\\
        &Frontal Sup  \\
        &Front Mid    \\
        &Supp Motor Ar\\
        &Front Sup Med\\
        &Front Med Orb\\
        &Insula       \\
        &Cingulum Ant \\
        &Cingulum Mid \\
        &SupraMarginal (L)\\
        &Precuneus    \\
        &Tempr Pol Sup\\
        &Caudate (R)\\
        \hline
        
        \multirow{5}{*}{Frontoparietal (FPN)} & Front Mid\\
        &Front Inf Ope\\
        &Front Inf Tri\\
        &Parietal Inf \\
        &Angular      \\
        \hline
        
        \multirow{5}{*}{Default Mode (DMN)} & Front Sup Med\\
        &Cingulum Ant \\
        &Cingulum Post\\
        &Angular      \\
        &Precuneus    \\
        \hline
        
        \multirow{7}{*}{Salience (Sal)} & Front Sup Orb\\
        &Front Mid    \\
        &Front Mid Orb\\
        &Front Inf Orb\\
        &Supp Motor Ar\\
        &Insula       \\
        &Cingulum Ant \\
        \hline
    
        \end{tabular}
    \end{adjustbox}
\end{table}

\newpage
\subsection*{Supplementary Subsection 7. Tractography.}
For the sake of completeness, in the following we describe in detail the methods used by Ref.~\cite{Deco2018} to obtain the human connectome. We clarify that we did not produce these connectome data sets, instead, we used data kindly provided by Gustavo Deco.

The structural connectivity between the 90 AAL regions was obtained from averaging across 16 healthy young adults (5 females, mean ± SD age: 24.75 ± 2.54). The linear registration tool from the FSL toolbox (www. fmrib.ox.ac.uk/fsl, FMRIB, Oxford) was used to coregister the EPI image to the T1-weighted structural image. The T1-weighted image was co-registered to the T1 template of ICBM152 in MNI space. The resulting transformations were concatenated and inversed and further applied to warp the AAL template from MNI space to the EPI native space, where interpolation using nearest-neighbor method ensured that the discrete labeling values were preserved. Thus the brain parcellations were conducted in each individual’s native space. The structural connectivity (SC) maps were generated for each participant using the dMRI data acquired. The two data sets acquired have different phase encoding to optimize signal in difficult regions. The construction of these structural connectivity maps consisted of a three-step process. First, the regions of the whole-brain network were defined using the AAL template as used in the functional MRI data. Second, the connections between nodes in the whole-brain network (i.e., edges) were estimated using probabilistic tractography. Third, data was averaged across participants.

The FSL diffusion toolbox (Fdt) was used to carry out the various processing stages of the diffusion MRI data using the 
default parameters of this imaging pre-processing pipeline on all participants. Following this preprocessing, we estimated the local probability distribution of fiber direction at each voxel. The probtrackx tool in Fdt was used to provide automatic estimation of crossing fibers within each voxel. 

The connectivity probability from a seed voxel \textit{i} to another voxel \textit{j} was defined by the proportion of fibers passing through
voxel \textit{i} that reach voxel \textit{
j} using a sampling of 5000 streamlines per voxel. This was extended from the voxel level to the region level, i.e., in an AAL parcel consisting of n voxels, 5000x n fibers were sampled. The connectivity probability $P_{ij}$ from region \textit{i} to region \textit{j} is calculated as the number of sampled fibers in region \textit{i} that connect the two regions divided by 5000 x n, where n is the number of voxels in region \textit{i}. The SC matrix was thresholded at 0.1\%, i.e., five streamlines.

For each brain region, the connectivity probability to each of the other 89 regions within the AAL was calculated. Due to the dependence of tractography on the seeding location, the probability from \textit{i} to \textit{j} is not necessarily equivalent to that from \textit{j} to \textit{i}. However, these two probabilities are highly correlated across the brain for all participants (the least Pearson \textit{r} = 0.70, p $< 10^-{50}$). As directionality of connections cannot be determined based on diffusion MRI, the unidirectional connectivity probability $P_{ij}$ between regions \textit{i} and \textit{j} was defined by averaging these two connectivity probabilities. This unidirectional connectivity was considered as a measure of the structural connectivity between the two areas, with ${C_{ij} = C_{ji}}$. The regional connectivity probability was calculated using in-house Perl scripts. For both phase encoding directions, 90x90 symmetric weighted networks were constructed based on the AAL90 parcellation, and normalized by the number of voxels in each AAL region; thus representing the structural connectivity network organization of the brain.

\newpage
\bibliography{dmf_5ht2a_ref}